# PHOTOPROCESSES ON THE $^6$Li NUCLEUS IN THE POTENTIAL CLUSTER MODEL

S. B. Dubovichenko, A. V. Dzhazairov-Kakhramanov

The photodisintegration of the $^6$Li nucleus through the channels $^3$He$^3$H, $^4$He$^2$H and the corresponding radiative capture are considered in simple two-cluster models. Potentials with forbidden states which reproduce the phase shifts of elastic scattering up to energies of 20 MeV are was used to describe cluster interactions. It is shown that such the potential cluster model provides an accurate description of total cross sections for photonuclear processes over the entire energy region under consideration.

Using simple potentials two-cluster models with intercluster interactions which include forbidden states (FS) [1], we recently calculated [2] the total cross sections for photodisintegration and radiative capture in the $^4$He$^3$H and $^4$He$^2$H system. These potentials describe the phase shifts of low-energy scattering and the forbidden states (FS) enable to take into account effectively the Pauli exclusion principle in cluster interactions [3]. We showed in [1] that some properties of $^6$Li and $^7$Li, which have relatively high probabilities of clustering in the aforementioned channels, can be described using such potentials. These potentials can be used to describe the basic properties of the ground states (GS) because the states in these systems have specific orbital symmetry [3].

The calculations in [1] were based on the potential cluster model. It is assumed that a nucleus consists of two structureless fragments whose properties coincide with the properties of the corresponding free particles. The wave function (WF) of the system is not antisymmetrized, but intercluster interactions include FSs. For this reason, the WF of the relative motion of the clusters oscillates in the interior of the nucleus, and scattering phase shifts satisfy the generalized Levinson theorem, are positive over the entire energy region, and tend to zero at high energies [3]. Using such a simple potential one-channel cluster model whose parameters were determined by fitting experimental data on scattering phase shifts and were not varied in the calculations, we obtained an accurate description of the total cross sections for photonuclear processes and of the properties of cluster nuclei [1, 2].

Recently the total cross sections for photonuclear processes in the $^4$He$^2$H and $^4$He$^3$H systems were calculated in [4] in the microscopic potential model and in [5] by the resonating group method (RGM). Using interactions involving FSs and three-body WFs GSs [6], calculated the total cross sections for the disintegration of the $^6$Li nucleus through the $^4$He$^2$H channel.

To calculate cross sections for radiative capture, we use the expression [2,4,5]

$$\sigma_c(J) = \frac{8\pi K^{2J+1}}{\hbar^2 q} \frac{\mu}{(2S_1+1)(2S_2+1)} \frac{J+1}{J[(2J+1)!!]^2} \sum_{m, m_i, m_r} |M_{Jm}(N)|^2, \tag{1}$$

where N = E or M, $M_{Jm}$ is matrix element for multipole operators in the cluster model, J is the multipole order; q is the wave number of the relative motion of clusters; $m_0$ is the nucleon mass; μ is the nucleus reduced mass; $S_i$ is the spins; K is the photon wave number. The photodisintegration cross section can be obtained from the capture cross section by using the detailed-balance principle [2,4,5].

The potential in the Gaussian expression with point coulomb term used to describe intercluster interaction

$$V(r) = -V_0 \exp(-\alpha r^2) + V_c(r)$$

(2)

he potential parameters are chosen in such a way as to ensure an accurate description of the partial scattering phase shift in the energy range up to 20 MeV by using a single set of parameters. Obtained parameters of potentials are listed in Table. and in [1]. The binding energy in the cluster channel, the charge radius, the probability of the radiative E2 transition $1^+ \rightarrow 3^+$, the elastic and inelastic Coulomb form factors, and the momentum distributions were described well in [1] on the basis of the potentials used.

| | $^6Li(^4He^2H)$ | | | | | | $^6Li(^3He^3H)$ | | |
|---|---|---|---|---|---|---|---|---|---|
| $L_J$ | $V_0$, | α, | $E_{FS}$, | S=1 | | | S=0 | | |
| | MeV | Fm$^{-2}$ | MeV | $V_0$, MeV | α, Fm$^{-2}$ | $E_{FS}$, MeV | $V_0$, MeV | α, Fm$^{-2}$ | $E_{FS}$, MeV |
| S | 76.12 | 0.2 | -33.2 | 90 | 0.18 | -47.0 | 85.0 | 0.18 | -43.3; -8.0 |
| $P_0$ | 68.0 | 0.22 | -7.0 | 52.5 | 0.2 | -4.9 | | | |
| $P_1$ | 79.0 | 0.22 | -11.7 | 65.0 | 0.2 | -9.8 | 74.0 | 0.2 | -14.1 |
| $P_2$ | 85.0 | 0.22 | -14.5 | 80.0 | 0.2 | -17.1 | | | |
| $D_1$ | 63.0 | 0.19 | --- | 72.0 | 0.18 | --- | | | |
| $D_2$ | 69.0 | 0.19 | --- | 85.0 | 0.18 | --- | 85.0 | 0.18 | --- |
| $D_3$ | 80.88 | 0.19 | --- | 90.0 | 0.18 | --- | | | |

Figure 1a shows the scattering phase shifts in the $^4He^2H$ system as functions of the c.m.s. energy. Experimental data on the phase shifts are presented in [8]. To obtain a phase shift value of $90^0$ at the experimental resonance energy 0.711(3) MeV [7], the interaction in the $D_3$ wave was some changed in [2] relative to the value used in [1]. In this case, the calculated width of the $3^+$ level equals 24(2) keV, and the measured value is 26 keV [7].

The contribution from $Q_{Jm}(S)$ to the total cross section is usually supposed to be small for E1 transitions. However, the spin term may be important for the $^4He^2H$ system because the cross section due to the orbital operator is small. It is well known that E1 transitions with ΔT = 0 are strongly suppressed in nuclei with M = 2Z. In the cluster model, this fact manifests itself in the factor $(Z_1/M_1 - Z_2/M_2)$, which is equal to zero in this case. An E1 transition due to the term $Q_{Jm}(L)$ can oc-cur in such nuclei only as the result of deviations from the condition M = 2Z, for example, the masses of the deuteron and the $^4$He nucleus are not integer: $M_d$ = 2.0135537 and $M\alpha$ = 4.0015061.

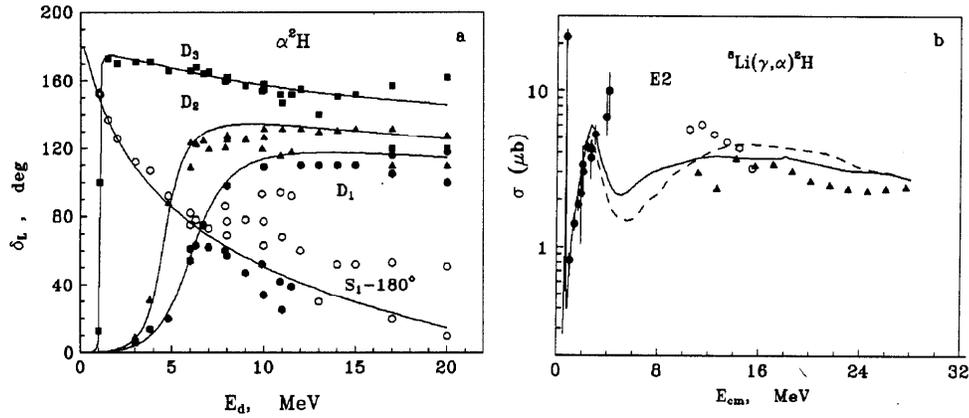

Fig. 1. (a) - Phase shifts for elastic scattering in the $^4$He$^2$H system. (b) - Total cross sections for $^6$Li photodisintegration through the $^4$He$^2$H channel.

Figure 1b shows that the calculated E2 cross section for disintegration reproduce the first experimental maximum which correspond to $D_3$ wave and has a second maximum at an energy of 2.5 - 3 MeV, in agreement with the experimental data also displayed in the Figure. This maximum corresponds to the resonance in the $D_2$ wave at an energy of 2.84 MeV. The spectrum of the nucleus contains and $D_1$ level at an energy of 4.18 MeV. However, the calculated cross section have not resonance behavior in the neighborhood of 4 MeV. At the same time, the corresponding phase shift has a resonance character. For comparison, the dashed curve in Fig. 1b shows the cross section calculated in [6] on the basis of three-body GSs WFs. Experimental data on the capture and disintegration are presented in [9].

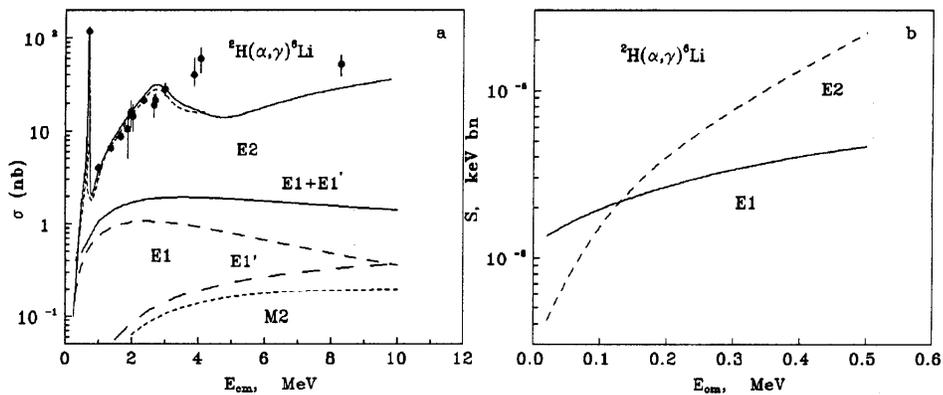

Fig. 2. (a) - Total cross sections for radiative capture from the $^4$He$^2$H system to the $^6$Li GS. (b) - Low-energy astrophysical S factors for radiative capture from the $^4$He$^2$H channel.

The solid line on Fig. 2a shows result our calculations of E2 the capture process from the D scattering states and the dotted curve displays the E2 cross section cal-culated on the basis of the potential model as presented in [4]. Figure 2a also displays calculated the cross sections for M2 capture from the P scattering states to the GS. The points show experimental data obtained in [9].

The terms of the cross section that are associated with the orbital (E1) and spin (E1') components were determined separately, and their interference was taken into account in the total cross section. Figure 2a shows that the cross section for the E1 process is considerably smaller than the cross section for E2 capture and that it makes virtually no contribution to the total cross section at energies above 0.2 MeV.

Figure 2b shows the S factors obtained from the cross sections for the E1 and E2 processes. It is shown, that the E1 process dominates at low energies. Linear extrapolation of the calculated S factors gives the values about $S(E2) = 3 \cdot 10^{-7}$ keV b and $S(E1) = 1.2 \cdot 10^{-6}$ keV b at zero energy. Thus, the total S factor is approximately equal to $1.5 \cdot 10^{-6}$ keV b.

Figure 3 shows the phase shifts calculated using these potentials, the experimental data reported in [10] (closed circles, triangles, and open circles), and the phase shifts calculated in the RGM [11] (crosses). To calculate the F wave phase shifts, the nuclear P wave potential with L = 3 was used (see Fig. 3b). In contrast to [1], spin-orbit interaction in the P, D, and F waves is taken into account in this study. Using the parameters of the S wave potential from [2], we were able to describe the RGM phase shifts (the dotted curve in Fig. 3a), but failed to reproduce the correct binding energy in the $^3$He$^3$H channel. To obtain the correct value -15.8 MeV [7] of the GS energy, we had to increase the potential-well depth to 105 MeV. The solid curve in Fig. 3a shows the phase shifts corresponding to this S wave potential. The use of a potential with such parameters leads to an incorrect value of the $D_3$ level energy. The correct value of -13.6 MeV relative to the threshold for the $^3$He$^3$H channel can be obtained using the parameters $V_0 = 107.5$ MeV and $\alpha = 0.18$ fm$^{-2}$ (the solid curve in Fig. 3a). The measured D wave phase shifts can be described only with the potential parameters presented in Table. The dotted curves in Fig. 3a show the phase shifts corresponding to these potentials.

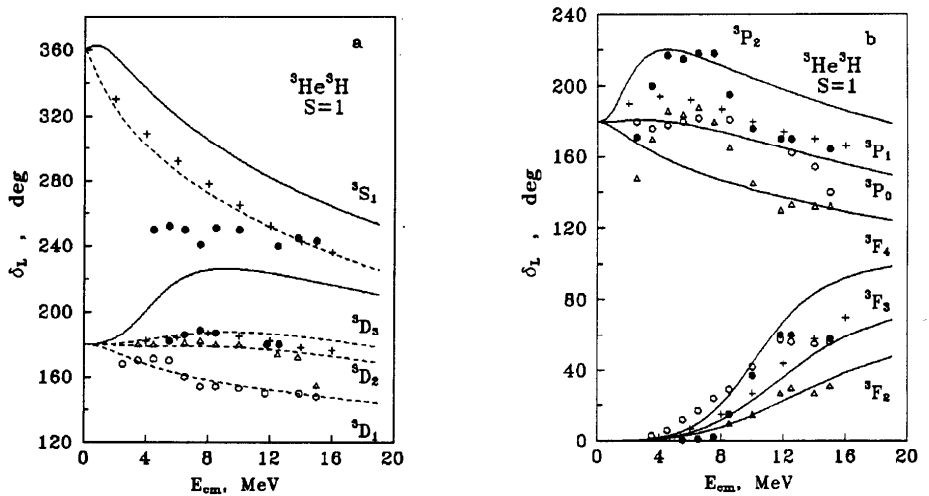

Fig. 3. Triplet phase shifts for elastic scattering in the $^3$He$^3$H system.

Figure 4a shows the total cross section for capture calculated on the basis of the changed S wave potential, along with the experimental data reported in [12]. It can be seen that calculations with such a potential and P wave interaction, which makes it possible to reproduce the energy dependence of the scattering phase shifts (see Fig. 3b), describe the experimental data well. We note that the experimental data presented in [13] differ noticeably from those presented in Fig. 4a.

We also considered an M1 transition from the spin singlet state to the spin triplet GS and E2 transitions from the states with L = 2 and $J_i$ = 1, 2, and 3 to the GS. The dotted and dashed curves in Fig. 4a show the calculated cross sections for the M1 and E2 transitions, respectively.

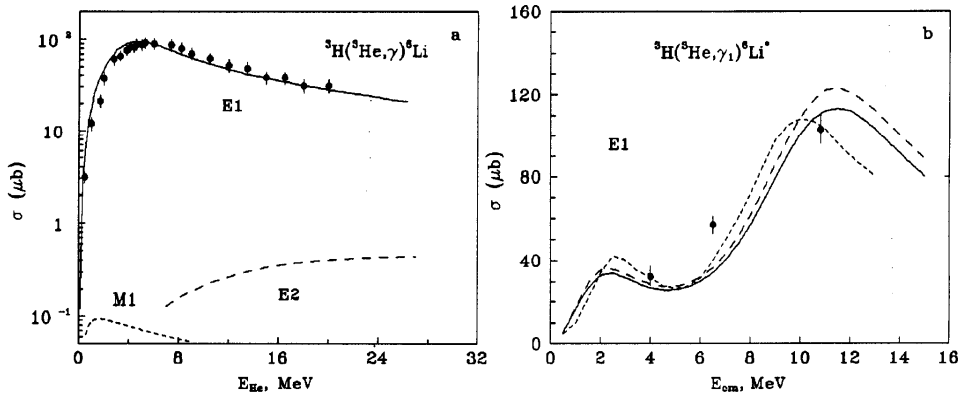

Fig. 4. (a) - Radiative capture total cross sections for the $^3$He$^3$H system to the $^6$Li GS and (b) - radiative capture from the $^3$He$^3$H system to the excited $3^+$ state of $^6$Li.

Figure 4b displays the calculated cross sections for capture to the $3^+$ level, experimental data (points), and theoretical results [14] (the dotted curve). The solid and dashed curves in this figure represent the results of calculations made with $D_3$ wave potentials of depths 105 MeV and 107.5 MeV, respectively.

The S factor for capture in the $^3$He$^3$H channel was calculated also. Linear extrapolation of the S factor gives a value of 0.06 keV b for the E1 process at zero energy.

In summary, the proposed cluster model with potentials that include FSs and which are consistent with the phase shifts for elastic scattering practically correctly describes the total cross sections for photodisintegration and radiative capture over the entire energy region in which experimental data are available. The successful application of the simple one-channel $^4$He$^2$H cluster model is due to the high probability of clustering in lithium nuclei. Not only the static properties, but also the form factors [1] and total cross sections for photonuclear processes can be described using a single set of potentials in the $^4$He$^2$H channel.

The experimental photo cross sections are also can be reproduced correctly in the $^3$He$^3$H channel. However it is not possible to make an agreement between the GS potential and the potential describing the scattering phase shifts. Only the use of an interaction with parameters reproducing GS characteristics is necessary to obtain the correct results.